\documentclass[prd,preprint,
               nofootinbib,amsmath,amssymb,
               tightenlines,floatfix]{revtex4}
\input epsf
\usepackage{bm}%        bold math
\def\be{\begin{equation}}
\def\ee{\end{equation}}
\def\bea{\begin{eqnarray}}
\def\eea{\end{eqnarray}}
\def\Eq#1{Eq.~(\ref{#1})}
\def\MM{{\mathcal{M}}}
\def\pres{\mathcal P}
\def\prob{P}
\def\TT{\mathcal{T}}

\advance\parskip 2pt

\renewcommand{\vec}[1]{{\bf #1}}
\newcommand{\vwall}{v}

\def\centerbox#1#2{\centerline{\epsfxsize=#1\textwidth\epsfbox{#2}}}

\begin{document}

\title
    {
      Electroweak Bubble Wall Speed Limit
    }

\author{Dietrich B\"odeker}
\affiliation{
Fakult\"at f\"ur Physik,
Universit\"at Bielefeld,
33501 Bielefeld,
Germany
}

\author{Guy D.~Moore}
\affiliation
    {%
      Institut f\"ur Kernphysik, Technische Universit\"at Darmstadt,
      Schlossgartenstra{\ss}e 2,
      64289 Darmstadt, Germany
    }%

\date {March 2017}

\vspace{10mm}

\begin {abstract}%
  {%
    In extensions of the Standard Model with extra scalars, the
    electroweak phase transition can be very strong, and the bubble
    walls can be highly relativistic.
    We revisit our previous argument that electroweak bubble walls can
    ``run away,'' that is, achieve extreme ultrarelativistic velocities
    $\gamma \sim 10^{14}$.  We show that, when particles cross the
    bubble wall, they can emit transition radiation.
    Wall-frame soft processes, though suppressed by a power of the
    coupling $\alpha$, have a significance enhanced by the
    $\gamma$-factor of the wall, limiting wall velocities to
    $\gamma \sim 1/\alpha$.  Though the bubble walls can move at
    almost the speed of light, they carry an infinitesimal share of
    the plasma's energy.
}
\end {abstract}

\maketitle

\section {Introduction}
\label{sec:intro}

A first order electroweak phase transition in the expanding Universe
proceeds through the expansion of bubbles into the high temperature
phase \cite{Kirzhnits:1976ts,Dine:1992wr}.  Potentially observable
consequences of this phase transition depend critically on the bubble
wall velocity $ v  $
\cite{Cohen:1993nk,Cohen:1994ss,Joyce:1994zt,Huber:2000mg}.  It is in
general quite difficult to compute $ \vwall $.  Most existing
microscopic calculations are for the case that the transition occurs due to
radiative corrections \cite{Turok:1992jp,Dine:1992wr,%
  Liu:1992tn,Khlebnikov:1992bx,Arnold:1993wc,Moore:1995si,%
  John:2000zq,Moore:2000wx}.  Generally these studies find that the
bubble wall velocity is less than the speed of sound.  However, in the
presence of singlet scalars, the electroweak phase transition can be
very strong, with a mean-field analysis already predicting a
first-order transition \cite{Huber:2000mg}.  In \cite{Bodeker:2009qy}
we argued that, in this case, the bubble wall velocity typically
becomes large, with the gamma factor $ \gamma \equiv (1-v^2)^{-1/2}
\gg 1 $. We presented a leading-order calculation of the ``friction''
(backwards force) on an advancing bubble wall, and showed that the
calculation simplifies tremendously in the $\gamma \gg 1$ limit, and
that the ``friction'' on the bubble wall approaches a constant.  This
occurs because of two compensating effects.  First, the density of
particles (as viewed in the wall's rest frame) rises due to Lorentz
contraction of the plasma, $n\propto \gamma$.  But the retarding force
from each particle diminishes as $\gamma^{-1}$, leading to a finite
$\gamma \to \infty$ limit.  This limit could be expressed in terms of
a modified effective potential, where the thermal part of
$V_{\mathrm{eff}}$ is replaced by a certain mean-field approximation.
This gives a simple criterion for whether the wall would speed up even
further (`run away') or not: if the phase transition is first order
with the mean field potential, then the wall will generally run away,
with the $\gamma$-factor growing linearly with propagation distance.

The difference between nonrelativistic and relativistic bubble walls
is important to baryogenesis.  The difference between relativistic
bubble walls and true runaway walls is relevant for possible
gravitational wave signatures of the transition
\cite{Espinosa:2010hh,Leitao:2015fmj,Katz:2016adq}, and deserves more
careful investigation.  The analysis of \cite{Bodeker:2009qy}
considered only the leading order in
the couplings, which means that it neglected any interactions of medium
excitations other than those with the electroweak bubble wall.  In
particular, we did not take into account that
particles may radiate (split) when they hit the wall.%
\footnote{%
  In footnote 8 of \cite{Bodeker:2009qy} it was claimed that particle
  splitting does not significantly modify the conclusions of the main
  text.  We will find that this is not true.}
Such radiation can occur because crossing the bubble wall causes
particles to accelerate, as their masses change from one side to the
other.  Furthermore, even if a particle remains massless, it can
radiate if it interacts with a particle (such as the $W$ boson) with a
phase-dependent mass, or if an interaction vertex changes in strength
from one phase to the other (for instance, a 3-point scalar vertex
proportional to the vacuum value of a field).  In these cases,
the ``radiation cloud'' surrounding (``dressing'') the particle
is not the correct cloud for the other phase, and the difference is
emitted as radiation.

This is the same phenomenon as transition
radiation:  when an electron moves from one medium into another medium
with a different index of refraction, the mismatch in the
electromagnetic ``cloud'' it should carry in each medium causes a
radiation associated with the transition between media -- transition
radiation \cite{Jackson:1998nia}.

In the current context, the emission of transition radiation induces a
force on the bubble wall.  The force depends on the momentum of the
emitted particle.  We will show that this force is dominated by
radiated particles which are soft in the wall frame.  We will show
that the probability to emit a gauge boson with a phase-dependent
mass, and the emitted spectrum of soft gauge bosons, approach a finite
$\gamma \gg 1$ limit.  Therefore the force on the wall \textsl{per
  particle}, arising from transition radiation, approaches a constant
in the large $\gamma$ limit.  Since the density of particles rises as
$n \propto \gamma$, the friction rises linearly in $\gamma$.
Therefore some $\gamma$ is sufficient to resist the forward pressure
driving the wall's acceleration.  Typically this occurs for $\gamma
\sim 1/\alpha$.

The remaining sections of the paper fill in the details in the
argument presented above.  The next section sets the stage with a
review of bubble wall kinematics.  Then Section \ref{sec:transition}
explains how to compute the rate of transition radiation for a bubble
wall.  We follow this with an outline of the rate of hard emissions in
\ref{sec:generic} and of the complications associated with soft
emissions in \ref{sec:IR}.  We end with a short conclusion.

\section{Review of bubble-wall kinematics}
\label{sec:review}

\begin{figure}[ht]
 \centerbox{0.6}{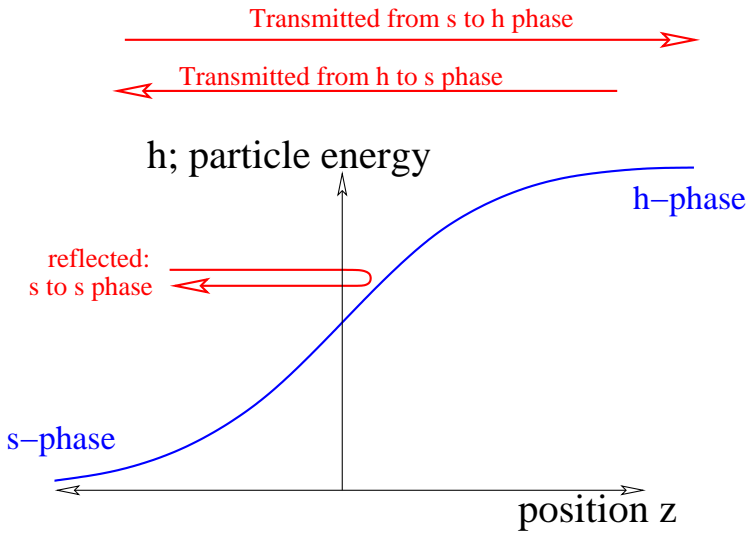}
  \caption{\label{fig1}
    Bubble wall geometry, and particle trajectories depending on    
    energy and direction.
  }
\end{figure}

We start with a lightning review of the main ideas in our first paper
\cite{Bodeker:2009qy}.  Consider an electroweak phase interface
(bubble wall) moving through the plasma.  This means that there is a
scalar field with a spacetime varying expectation value
$\varphi(t,\vec r)$.  Also there will be one or more particles whose
masses arise from this expectation value, $m \propto \varphi(t, \vec
r)$.  We assume that the wall reaches a steady-state velocity $v$ and
gamma-factor $\gamma = 1/\sqrt{1-v^2} \gg 1$ relative to the plasma in
front of it, which we will try to determine self-consistently.  Since
bubbles are macroscopic, we can take the wall to be planar and choose
coordinates where $\varphi = \varphi(t,z)$; in the wall's rest frame
$\varphi = \varphi(z)$, and $v$ is the velocity of the plasma
approaching the wall from in front.  The geometry is summarized in
Fig.~\ref{fig1}.

The wall is pushed forward (to the left in the figure) by the vacuum
energy difference between phases.  There is a restraining force from
particles, which must balance the forward force for the wall to reach
a steady state.  Therefore it is essential to compute this restraining
force, which we do by computing the impulse from a single particle
and integrating over the flux of particles.  By assumption the wall is
moving very fast relative to the plasma, $\gamma \gg 1$, so almost all
particles have an energy $E \sim \gamma T$ much larger than the mass
change in crossing the wall.  Furthermore the particles have much
shorter wavelength than the wall's thickness, so the WKB approximation
is good and we can neglect reflection.  Another simplification due to
$\gamma \gg 1$ is that particles only approach the wall from the
symmetric-phase side.

We will choose coordinates such that the Higgs field varies along the
$z$ axis from a large ``h'' value at positive $z$ to a small ``s''
value at negative $z$.  We work in the wall rest frame, meaning the
frame obtained from the s-phase plasma rest frame by a purely
$z$-directed boost.  Consider a particle $ a $ of energy $E_a \sim
\gamma T $ in the s phase impinging on the wall from the left.  Our
frame is time and $\vec r_\perp$ independent, so $E$ and $\vec
p_\perp$ are conserved.  But the wall breaks $z$-translation
invariance, so $p_z$ is not conserved.  Any change to a particle's
$p_z$ must be supplied by an impulse from the bubble wall; so we need
the particle's $p_z$-change in crossing the wall.  Before hitting the
wall, the energy is related to transverse momentum $\vec p_\perp$ and
$z$-momentum $p_z$ as $E^2 = p_z^2 + p_\perp^2 + m_{\rm s}^2$, where
the subscript s refers to the symmetric phase and $ p _\perp \equiv |
\vec p _\perp | $.  Since $E \sim \gamma T$ and $p_\perp^2 \sim T^2$,
we can expand in $E,p_z \gg p_\perp, m$, in which case the momentum
transferred to the wall is
\be  
  \Delta p _ { 1 \to 1 } = p_{z,\rm in} -
p_{z,\rm out} \simeq \frac{ m ^ 2 _ { a,\mathrm h } - m ^ 2 _ {
    a,\mathrm s } } { 2 E} \,.
    \label{dp1}
\ee
This should be integrated over the phase space and occupancy of
particles impinging on the wall.  Because the reflection coefficient
is approximately zero and no particles hit the wall from the
h-phase, no ``news'' of the wall's approach has reached the
s-phase, and the s-phase occupancies take their equilibrium
values.  So the pressure on the wall is
\begin{align} 
\label{DeltaP}
\pres_{1\to 1} &= \sum_a \nu _a\int \frac{d^3 p}{(2\pi)^3} f_a(p) \times
\frac{m^2_{a,\rm h}-m^2_{a,\rm s}}{2E}
\nonumber \\
&=
\sum_a \nu_a \int \frac{d^3 p}{(2\pi)^3 2E} f_a(p)
\; \Big(m^2_{a,\rm h} - m^2_{a,\rm s} \Big) \,.
\end{align} 
Here the sum $\sum_a$ is over all particle types, and $\nu_a$ is the
degeneracy (number of spin and color or other group labels).  The
notation $1\to 1$ means that we are considering a single particle
approaching the wall, remaining a single particle after crossing the
wall.  The combination $d^3 p/E$ is frame-independent, leading to a
$\gamma$ independent pressure (in the large-$\gamma$ limit, used to
reach \Eq{DeltaP}).
This result -- that the pressure restraining the bubble wall
approaches an asymptotic value rather than growing with increasing
$\gamma$ -- was the main result of our previous paper \cite{Bodeker:2009qy}.

\section{Transition Splitting}
\label{sec:transition}

Transition radiation, or transition splitting, can occur if an
initial particle species $a$ couples to final species $b,c$ and
either the strength of their interaction, or one or more particle
mass, differs between the two phases.
The best known case is the emission of light by a charged particle,
say an electron, via the process $e^- \to e^- \gamma$, when crossing
an interface which changes the photon dispersion.%
\footnote{This is distinct from Cherenkov radiation.  In particular,
  transition radiation occurs even if the electron velocity is below the
  Cherenkov velocity in each medium.}
Several species change mass in going from the symmetric to the Higgs
phase, and they all have gauge and/or Yukawa interactions allowing
$a\to bc$ type processes; so transition radiation can certainly occur
when a particle hits the electroweak interface.  Such a splitting
changes the kinematics and increases the $p_z$ transferred to the
wall.  Some of the incident particle's energy goes
into the mass of the additional particle, and
some into the relative transverse momentum of $ b $ and $ c $, and
there is less energy left for the longitudinal momenta of
the produced particles.
So we need to consider these processes to see whether they have
an important effect on the previous arguments.

To compute the force due to transition radiation, we have to
understand the flux of incident particles, the probability for each to
undergo transition radiation, and the impulse to the wall if it does.
We start with the flux of particles, which is simple.
In the wall's rest frame, the density of particles in front of the
wall, and the flux impinging on the wall, are
\begin{equation}
  \label{densityflux}
  \mbox{density } = \sum_a \nu_a \int \frac{d^3 p}{(2\pi)^3} f_a(p) \,,
  \qquad
  \mbox{flux } = \sum_a \nu_a \int \frac{d^3 p}{(2\pi)^3} \frac{p_z}{p^0}
  f_a(p) \,.
\end{equation}
The backwards pressure due to splittings is found by inserting the
differential probability to split, times the momentum transferred by the
splitting process, inside this integrand:
\begin{equation}
\label{P_v1}
  \pres_{1\to 2} = \sum_a \nu_a \int \frac{p_z d^3 p}{p^0 (2\pi)^3} f_a(p)
  \;\times \sum_{bc} \int dP_{a\to bc} \: \times \:
  ( p_{z, \rm s}-k_{z,\rm h}-q_{z,\rm h}) 
  \,.
\end{equation}
Here $k$ and $q$ are the final momenta of the $b$ and $c$ species
respectively.  The notation $1\to 2$ means we are computing the
pressure arising when 1 incoming particle becomes 2 final particles
after crossing the wall.

To be more precise about the differential splitting probability
$dP_{a\to bc}$, consider a single
particle impinging on the wall.  To work with a properly normalized
state, we integrate the (improperly normalized) momentum-space states,
\begin{equation}
\label{norm}
  \langle \vec p' | \vec p \rangle = 2p^0 (2\pi)^3 \delta^3(\vec p- \vec p') 
  \,,
\end{equation}
over a wave packet which builds a properly normalized single particle
initial state,
\begin{equation}
\label{wavepacket}
  | \phi    \rangle \equiv \int \frac{d^3 p'}{(2\pi)^3 2p_0'}
  \phi(\vec p') |\,\vec p'\rangle \qquad \mbox{with} \qquad
  \int \frac{d^3 p}{(2\pi)^3 2p_0} |  \phi(\vec p) | ^ 2 = 1 
  \,.
\end{equation}
We are labeling momentum space states in terms of their incoming
symmetric-phase momentum; because there is a bubble wall, they
actually vary nontrivially  in $ z $-direction, 
\begin{equation}
  \label{xspace}
    \langle \vec r | \vec p \rangle 
   = 
   \sqrt{2p^0}  e ^{ i \vec p _\perp \cdot \vec r _\perp } \chi_  p(z) 
   \quad \mbox{(for scalars)}
\end{equation}
with $\chi_p(z) = \exp ( ip _ z z ) $
in the s phase, neglecting any
reflected wave; but the behavior of $\chi_  p$ near the wall and in
the broken phase must be found by explicitly solving the associated
free particle evolution equation in the presence of the bubble wall.
For spinor or vector fields, one should replace $ \chi  _  p$ with
an appropriate spinor or vector solution to the linearized 
Dirac or Yang-Mills equation in the wall background.

An integral over the final state momenta, with momentum-basis states,
does constitute a properly normalized treatment of the final states.%
\footnote{With the normalization of \Eq{norm},
  $(2\pi)^{ -3 } \int   \, d^3 k ( 2k^0 ) ^{ -1 } 
| \vec k \rangle \langle \vec k | $ is a
  properly normalized projection operator.}
So the splitting probability alluded to above is the integral over
final state phase space of the squared $\cal T$-matrix element of the
initial state with the multiparticle final state;
\begin{align}
  \label{Psplit}
  \int d \prob_{a\to bc} & \equiv
  \int \frac{d^3 k \: d^3 q}{(2\pi)^3 2k^0 (2\pi)^3 2q^0}
%  \phi^*(\vec p_2) \phi( \vec p_1) \;
  \langle \phi % \vec p _ 2
  | \TT | \vec k \vec q \rangle
  \langle \vec k  \vec q | \TT | \phi % \vec p _ 1
  \rangle
  ,
\end{align} 
implicitly summing over final state spin and color indices.

The bubble wall is invariant in time and the transverse directions,
ensuring that energy and transverse momentum are conserved, so the
transition matrix element between momentum states is
\begin{align}
\label{MbarM}
   \langle \vec k  \vec q | \TT | \vec p \rangle
  & =
   \int d^4 x \; \langle \vec k \vec q | \mathcal{H}_{\mathrm{int}}
   | \vec p \rangle
   \nonumber \\ & =
   (2\pi)^3 \delta^2(\vec p_\perp {-} \vec k_\perp {-} \vec q_\perp)
  \delta  (p^0 {-} k^0 {-} q^0 )  \MM 
  \,,
\end{align}
with 
\begin{align}
  \MM & \equiv  \int\!\! dz \; \chi^*_ k(z) \chi^*_ q(z)  V(z) \chi_p(z) 
  \label{Mis} 
  \,.
\end{align}
Here $V(z)$ is the
contraction of the interaction Hamiltonian density
with all other state information (spinors and polarizations), which
would be the same as the interaction matrix element if we were
considering simple plane wave states.
Note that $V(z)$ has dimensions of energy, while $\MM$ is
dimensionless.

Applying \Eq{wavepacket} and \Eq{MbarM} to \Eq{Psplit}, we find
\begin{align}
  \label{Psplit2}
  \frac{p_z}{p} \int dP_{a\to bc}
  & =
  \int \frac{p_{1z} d^3 p_1 d^3 p_2}{(2\pi)^6 2(p_1^0)^2 2p_2^0}
  \phi^*(\vec p_1) \phi(\vec p_2) (2\pi)^3 
  \delta^2(\vec p_{1\perp} {-} \vec p_{2\perp})
  \delta (p_1^0 {-} p_2^0  ) 
  \nonumber \\
  & \times \int \frac{d^3 k d^3 q}{(2\pi)^3 2k^0 (2\pi)^3 2q^0}
  (2\pi)^3 \delta ^ 2 (\vec p_{1\perp}- \vec k_ \perp- \vec q_\perp) 
  \delta (p_1^0-k^0-q^0  )
  | \MM | ^ 2
  \,.
\end{align}
Note that $p_z \delta(p_1^0-p_2^0)/p^0 = \delta(p_{1z}-p_{2z})$, so
the $p_2$ integration can be performed using the delta functions,
leaving a factor of $1/2p^0$.  The remaining $\vec p_1$ integral over the
wave packets gives 1, and replaces $\vec p_1 \to \vec p$, the central value of
the wave packet, in the remaining expressions.  Also note that
$|\MM|^2$ in \Eq{Psplit2} is final state spin and color summed.

Inserting the resulting $dP_{a\to bc}$ into \Eq{P_v1}, and restoring
the final state Pauli blocking or Bose stimulation factors which we
neglected to write so far, we find that the backwards pressure on the
bubble wall from splitting processes is
\begin{align}
  \label{Pis}
  \pres_{1\to 2}
  = 
  \sum_{a,bc} \nu_a \int \frac{d^3 p}{(2\pi)^3 2p^0}
\int & \frac{d^3 k d^3 q }
     {(2\pi)^6 2k^0 2q^0}
     f_p [1{\pm}f_{k}][1{\pm} f_{q}] ( p_{z, \rm s}-k_{z,\rm h}-q_{z,\rm h} ) 
     \nonumber \\
     \times & (2\pi)^3 \delta^2(\vec p_\perp-\vec k_{\perp} - \vec q_{\perp})
     \delta(p^0-k^0-q^0) |\MM|^2 
     \,.
\end{align}

The matrix element $\MM$ can be nontrivial if either $V(z)$ or one of
the $\chi(z)$ is a nontrivial function of $z$.  If both $V$ and all
$\chi$ are $ z $-independent, it reduces to $\MM \propto
\delta(p_z-k_{z}-q_{z})$, indicating that no momentum is transferred
to the wall.  In this case the process is either kinematically
forbidden, or it represents a kinematically allowed decay $a\to bc$
which occurs anywhere in space, regardless of the wall.  Therefore the
interesting cases are when either the vertex, or one or more particle
masses, differ between phases.

\section{Semi-soft emission}
\label{sec:generic}

To make further progress, we will specialize to a kinematic case which
will be sufficient to determine where the dominant friction arises.
We will consider
\be
   \label{kin_approx}
   p_z \sim \gamma T \,, \qquad
   p_z \gg k_{z} \gg L^{-1},m
   \,,
\ee
that is, $k_{z}$ intermediate between the ``hard'' scale $p_z$ and
the soft scale set by masses.  Here $L^{-1}$ is the inverse wall
thickness, which is related to the scalar (Higgs) field mass.
We will also assume
\begin{align}
  \label{kperp}
  k_\perp^2 \sim m^2 \quad ( \ll k_z m \; \mbox{by
    Eq.(\ref{kin_approx})} \,), 
\end{align}
which will prove to be the most important $k_\perp$ range.
These kinematical approximations will simplify
things enough to get explicit expressions, which we can use to
determine the most relevant processes and momentum ranges.  We will thereby
find that the most important processes are those which emit vector bosons
with phase-dependent masses, and that the most important kinematic
range for this process is in fact ``soft'' $k$, with $k_\perp \sim k_z
\sim m$.  Our analysis will identify this momentum range as the most
important, but its quantitative evaluation then lies outside of the
range of validity of the analysis.

The $ z $-dependent longitidinal momenta can now be approximated as 
\be
\label{E_approx}
k_z(z) = \sqrt{k_0^2 - m^2(z) - k_\perp^2}
 \simeq k^0 - \frac{m^2(z) + k_\perp^2}{2k^0} 
 \,.
\ee
Using $ q ^ 0 \simeq p ^ 0$ the integral (\ref{Pis}) for the pressure
can be simplified to
\begin{align}
  \label{phasespace}
  \pres_{1\to 2} & = \sum_{a,b,c} \nu_a \int \frac{d^3 p}{(2\pi)^3 4p _ 0^2} f_p
  \int \frac{d^2 k_\perp}{(2\pi)^2} \int _ 0 ^ {\infty  } 
  \frac{dk ^ 0}{2 \pi   2 k  ^ 0}
       [1 {\pm} f_k][1 {\pm} f_{p-k}]
       \frac { k_\perp^2 + m_{b,\rm h}^2}{2k ^ 0} |\MM|^2
       .
\end{align}

Because we assume $p ^ 0,k ^ 0 ,q ^ 0 \gg L^{-1}$, we may treat the
mode functions $\chi (z)$ in the WKB approximation, 
\be
   \chi  _ k(z) 
   \simeq 
   \sqrt\frac{k_{z,\mathrm s}}{k_z(z)} 
   \exp \left ( i\int_0^z k_z(z') dz' \right ) 
   \quad \simeq \quad e^{ik^0 z} \exp \left( -\frac{i}{2k^0} \int_0^z
(m^2(z')+k_\perp^2) dz' \right) \,.
 \label{cwkb} 
\ee

Making a small transverse boost so $\vec p_\perp=0$, and
introducing the energy fraction 
\begin{align}
   x \equiv  k ^ 0/p ^ 0 \ll 1
   \label{x} 
\end{align} 
carried by particle $b$, we find
\begin{align}
  \chi_a(z) \chi_b^*(z) \chi_c^*(z) 
  &    =  
    \exp \left \{ 
    \frac{i}{2p^0}
    \int_0^z \left(
   m^2_a - \frac{m_b^2 +  k_\perp^2}{x} - \frac{m_c^2 +  k_\perp^2}{1-x}
   \right) 
   d z'
   \right \}
   \label{ccc} 
   \,.  
\end{align} 
Under our assumptions, this phase is small for
$|z|<L$ (inside the wall).  Therefore we can replace $\int dz' m^2$
with $-|z| m^2_\mathrm{s}$ for $z<0$ and $z m^2_\mathrm{h}$
for $z>0$.  In this case,
and assuming that $V$ takes a value $V_\mathrm s$ in the symmetric phase and
$V_\mathrm h$ in the Higgs phase, \Eq{Mis} becomes 
\be
   \MM = V _ {\rm s }
   \int\limits _ { - {\infty  } } ^ 0
   dz \exp \left [  iz \frac { A _ {\rm s} } { 2p ^ 0}  \right ]
   + V _ {\rm h} \int\limits _ 0 ^ {\infty} dz
   \exp \left  [ iz \frac { A _ {\rm h} } { 2p ^ 0 } \right ]
   =
   2i p ^ 0\left ( \frac{ V _ { \rm h } } { A _ { \rm h } }
     - \frac{ V _ { \rm s } } { A _ { \rm s } } \right )
 \,
   \label{mthin}
\ee
with
\be
-A  = \frac{k_\perp^2}{x(1-x)}
              -m_{a}^2 + \frac{m_{b}^2}{x}
              + \frac{m_{c}^2}{1-x}
              \simeq \frac{k_\perp^2 + m_b^2}{x} 
              \,.
               \label{ax} 
\ee
In the two cases of most interest we find
\begin{align}
  |\MM|^2 
  & \simeq
      4p _ 0^2 \dfrac{|V_{\rm h} - V_{\rm s}|^2}{A^2} 
      &
   \quad ( V_{\rm h} \neq V_{\rm s} ) 
   \,,
  \label{fixedA}
   \\
  |\MM|^2 
  & \simeq
    4p _ 0^2 |V|^2 \dfrac{(A_{\rm h}-A_{\rm s})^2}{A_{\rm h}^2
    A_{\rm s}^2} 
  &
  \quad ( A_{\rm h} \neq A_{\rm s} ) 
  .
  \label{fixedV}
\end{align}

Next we investigate the form of $V$, the $1\to 2$ scattering matrix
element.  For the cases with $|V|^2 \propto k_\perp^2$, it has a
simple relationship
to the standard DGLAP splitting kernel, with
$P_{b\leftarrow a}(x) = |V|^2 x(1-x) / 16\pi^2 k_\perp^2$.  The
additional factors arise from the phase space integration involved
in treating the $1\to 2$ process in the DGLAP setting.
We list the value of $|V|^2$ for the most interesting processes
in Table \ref{Vtable}.  In computing the expressions in the table we
have again expanded systematically in large
$p_z \gg k_{z} \gg m,k_\perp$.
We have also considered a simple group, so for instance for $W$, $ Z $ boson
emission from a doublet matter field one should use $C_2[R] = 3/4$ if
we neglect the hypercharge interaction.  To properly take into account
the mixing between SU(2) and U(1) interactions to find $Z$-boson
radiation at finite weak mixing angle requires a little more work.

\begin{table}
  \begin{center}
{\renewcommand{\arraystretch}{1.5}
  \begin{tabular}{|c|c|} \hline
    $\;\;a(p) \to b(k) c(p{-}k)\;\;$ &
    $\;\;|V^2|\;\;$ \\ \hline
    $S \to V_T S$ &                                   \\
    $F \to V_T F$ & $4 g^2 C_2[R] \dfrac{1}{x^2} k_\perp^2$  \\  
    $V \to V_T V$ &                                   \\ \hline
    $S \to V_L S$ &                                   \\
    $F \to V_L F$ & $4g^2 C_2[R] \dfrac{1}{x^2} m^2$  \\
    $V \to V_L V$ &   \\ \hline
    $ \vphantom{\Bigg [}$
    $F \to F V_T$ & $2g^2 C_2[R]\dfrac{1}{x} ( k_\perp^2 + m _ b ^ 2 ) $\\ \hline
    $ \vphantom{\Bigg [}$
    $V \to F F$ & $2g^2 T[R]\dfrac{1}{x} (k_\perp^2+m_b^2)$      \\ \hline
    $S \to S V_T$ & $4g^2 C_2[R] k_\perp^2$             \\ \hline
    $F \to S F $  & $y^2 (k_\perp^2 + 4m_a^2) $         \\ \hline
    $S \to S S $  & $\lambda^2 \varphi^2$              \\ \hline
  \end{tabular}}
  \end{center}
  \caption{\label{Vtable}
    Squared vertex functions for the most interesting transition
    radiation processes, averaged/summed over initial/final state
    spins and group indices.  Here $S$ is a scalar, $F$ is a spin 1/2
    fermion, and
    $V_T,V_L$ are transverse and longitudinal vector bosons.
    $C_2[R]$ is the second Casimir of the representation of the
    incoming particle, and $T[R]$ is the trace normalization (Dynkin
    index) of the final particle, and $\varphi$ is the phase-dependent
    background value of the scalar field.}
\end{table}

The vertices divide into two groups.  Some, such as those involving
longitudinal gauge bosons or the $S \to SS$ process, have a phase
dependent value, which however does not grow with $p ^ 0$ at fixed $p
^ 0/k ^ 0=1/x$ ratio.  Others scale with $k_\perp^2$ and therefore
increase as all momenta are raised simultaneously; but these terms are
the same in the two phases.  This makes sense: Any dependence of
$|V|^2$ on the phase must arise from an interaction with the scalar
expectation value, and should therefore be proportional to $m^2$.  The
only way for $| V | ^ 2$ to carry no powers of the expectation value
is for it to be built out of an invariant of the momenta, such as $p _
\mu k ^ \mu \sim ( p ^ 0 /k ^ 0 ) k_\perp^2$.  The quantity $p _ 0^2$
itself involves the dot of $p^\mu$ with the wall normal vector and
cannot appear without an $m^2$ factor.

The most small-$x$ singular terms are those involving the emission of
a soft vector boson, with $| V | ^ 2 \sim x^{-2}$.  All other terms scale
as $x^{-1}$ or less.  We will now see that this leads to soft vector
boson emission dominating the friction on the bubble wall, and that
the $x^{-2}$ behavior predicts that the friction is controlled by the
small-$x$ region.  To see this, insert the matrix element for
$F \to V_T F$ into \Eq{fixedV}, using \Eq{ax}:
\be
   \label{Msq_FVF}
   |\MM|^2 \simeq 4g^2 C_2[R] \frac{k_\perp^2}{x^2}
   \; \times \; 4p _ 0 ^2 \frac{m_ { V, \rm h }  ^4}{x^2} \; \times \;
   \frac{x^4}{k_\perp^4 (  k_\perp^2 + m_{ V, \rm h }  ) ^ 2}
   \,.
\ee
The overall dependence is as%
\footnote{%
  In deriving \Eq{Msq_FVF} we have neglected thermal masses.  This is
  a good approximation if the phase transition is strong so
  $m^2 \gg g^2 T^2$.  Including thermal masses raises some subtle
  issues, but it would be necessary in a careful quantitative
  treatment if the phase transition is not very strong.}
\be
   \label{Msq_simp}
   |\MM|^2 \simeq 16 g^2 C_2[R] \; p _ 0^2 \;
   \frac{ m_{V,\rm h}^4}{k_\perp^2
     (k_\perp^2 + m_{V,\rm h}^2)^2} \,.
\ee

The $\vec k_\perp$ integral in Eq.~(\ref{phasespace}) is now
dominated by $k_\perp^2 \sim m^2$, justifying \Eq{kperp} and giving
approximately
\be
\label{kperp_int}
   \int \frac{d^2 k_\perp}{(2\pi)^2} 
   \frac{1}{k_\perp^2(k_\perp^2+m _ { V,\rm h } ^2)}
   \sim \frac{1}{24\pi m^2} \,,
\ee
and the $p,k$ integrals are of the form
\be
\label{pint_kint}
\int \frac{d^3 p}{(2\pi)^3} f_p \; \times \;
m _ { V, \rm h } ^4 \int \frac{dk ^ 0 }{k _ 0 ^2} 
[1 {\pm} f_k][1 {\pm} f_{p{-}k}] \,.
\ee
The $p$ integral counts the wall-frame density of incoming particles,
and scales as $\gamma T^3$.  The factor $[1{\pm}f_{p{-}k}]$ is order
1, and the factor $[1{\pm} f_k]$ becomes 1 when we consider the
difference between emission and absorption processes.
The $k ^ 0$ integral is small-$k ^ 0$ divergent.
It should be  cut off by the mass scale, giving a pressure of form
\be
  \label{P_propto_gamma}
  \pres_{1\to 2}  \sim g^2 C _ 2 [R] n_p m \propto \gamma g^2 m T^3 \,.
\ee
The important feature is the scaling with the incoming particle
density, and therefore the linear proportionality to $\gamma$.

If we consider instead the emission of a longitudinal vector boson,
with $| V | ^ 2 \sim g^2 m_ { V , \rm h } ^2/x^2$, we find
\be
\label{M_VL}
|\MM|^2 \simeq 4p^2 \; \frac{g^2 C_2[R] m_{V,\rm h}^2}{x^2} \;
\frac{x^2}{(k_\perp^2 + m^2)^2} \,.
\ee
The $k_\perp^2$ integral is now only logarithmically small-$k_\perp$
dominated, but the result is parametrically the same if one neglects
logarithms.

However, Table \ref{Vtable} shows that only these processes give rise
to a $| V | ^ 2 \propto 1/x^2$ behavior. This is familiar from the DGLAP
kernels, where only soft vector emission has a $1/x$ enhancement.
Substituting a less $x$-singular expression, such as $| V | ^ 2$ for the
$F \to F V$ process, in \Eq{Msq_FVF}, we find that \Eq{pint_kint}
becomes
\be
\label{Pint_Kint2}
\int \frac{d^3 p}{(2\pi)^3 p} f_p \times m^2 \int \frac{dk ^ 0 }{k ^ 0} \ldots
\ee
which is only logarithmically small-$k$ singular.  More importantly,
the $p$ integration gives the invariant particle density
$\int d^3 p  f_p /p\propto T^2$, rather than the wall-frame density
$\int d^3 p f_p \propto \gamma T^3$.  The result is not enhanced at
large $\gamma$, or at most it is enhanced by logarithms of $\gamma$,
rather than a power.  Therefore such processes are subdominant.

Note also that the emission of a vector boson which is massless in
each phase (say, the gluon) also does not give rise to a
linear-in-$\gamma$ contribution.  Consider again the $F\to V_T F$
process, but with $m _ V ^2=0$ in each phase, and $m_F^2$ changing
between phases.  Then $A_{\rm h}-A_{\rm s} \sim m_F^2$ without a $1/x$
factor, making \Eq{Msq_FVF} less singular by $x^2$.

Therefore the friction from transition radiation is dominated by the
emission of vector bosons with phase-dependent masses, with backwards
pressure of parametric form $\pres_{1\to 2} \sim \gamma g^2 m_ V T^3$.

\section{Infrared behavior}
\label{sec:IR}

We have shown above that the friction from transition radiation is
dominated by gauge bosons receiving a mass in the transition, with the
most important momentum range being $k_\perp \sim k_z \sim m _ V
\equiv m$ in the wall frame.  Here we will outline the challenges
associated with a complete calculation.  However, we will not carry
out such a calculation, as it is technically complicated and sensitive
to the details of the (beyond the Standard Model) physics giving rise
to the transition.

Since the dominant emitted momentum is $k_z \sim m$, we must revisit
two approximations we made above.  First, in \Eq{mthin}, we
approximated the phase change in traversing the wall to be small.
This amounts to the approximation
\be
   \label{smallphase}
   \mbox{small phase change:} \qquad
   L \; \frac{k_\perp^2 + m^2}{k _ z} \ll 1\,, \qquad \mbox{or} \qquad
mL \ll 1 \quad \mbox{if $k_\perp,k _ z\sim m$}.
\ee
Second, in treating the wave functions of all particles in the WKB
approximation, we assumed
\be
\label{WKBcondition}
\mbox{WKB approximation:} \qquad
k_z L \gg 1\,, \qquad \mbox{or} \qquad
mL \gg 1 \quad \mbox{if $k_\perp,k_z \sim m$}.
\ee
Clearly, in the soft regime, one or the other of these approximations
will break down.

To see what kind of numerical factors can be involved
we do an explicit computation assuming that the wall is thick,
$ L m \gg 1 $.  Note that $L $ is of order an inverse
scalar mass, so this represents the case where the scalar mass is
smaller than the $W$-boson mass, which appears unlikely.  Nevertheless
we will present the calculation to illustrate how the IR cutoff can
occur.  We assume a tanh form for the bubble wall,
\begin{align}
   m^2 ( z )
   =  \frac 12 \left ( m_{\rm h}^2 +  m_{\rm s}^2 \right )
   +    \frac 12 \left ( m_{\rm h}^2 - m_{\rm s}^2 \right )
   \tanh ( z/L )
   \label{tanh}
\end{align}
and we assume $ z $-independent $ V = C g k _\perp /x$. The masses in \Eq{tanh} have
contributions from the Higgs expectation value $ h $, and also
$ h $-independent ones, which are proportional to $ T $.  As before,
we take $x \ll 1$, so that
\begin{align}
   { \MM }  & \simeq
   V \int d z \exp \left (
      i z \frac {  k _\perp ^ 2 } { 2 k  ^ 0}
     +  \frac i { 2 k ^ 0 } \int _ 0 ^ z d z' m^2 ( z' )
     \right )
     .
\label{Mexplicit}
\end{align}
This gives
\begin{align}
  { \MM }  & \simeq        \frac {  V L } 2
     \exp \left (   - i \log 2\,   \frac L   { 4 k } \Delta  m^2 \right )
%     \nonumber \\
%     & \quad \times
     \frac {
     \Gamma  \left ( -i  \dfrac L  { 4 k }  (  k _\perp ^ 2 + m _ {\rm h}
         ^ 2 )  \right )
       \Gamma  \left ( i  \dfrac L  { 4 k }  (  k _\perp ^ 2 + m _ {\rm s}
         ^ 2 )  \right )
       } {
               \Gamma  \left ( -i  \dfrac L  { 4 k }  \Delta   m^2  \right )
       }
     \label{Mexpl}
\end{align}
with $ \Delta  m ^ 2 \equiv m _ {\rm h } ^ 2 - m _ {\rm s } ^ 2 $. 
Note that for $ (L m)(m/k) \ll 1 $, $ k _ \perp \gg m$ this agrees
with (\ref{Msq_simp}), as it should.
Using  that
$\left | \Gamma   ( i y ) \right | ^ 2 =  \pi  / [ y \sinh ( \pi  y ) ]   $
for real $ y $ \cite{abramowitz+stegun} we obtain
\begin{align}
  | \MM | ^ 2 & =
  \frac { \pi  L | V | ^ 2 k \Delta  m^2 }
  { (  k _\perp ^ 2 + m^2_{\rm h} )
    ( k _\perp ^ 2 + m^2_{\rm s} )
    }
    \,
%    \nonumber \\
%    & \quad \times
    \frac {
    \sinh \left (  \dfrac { \pi  L  }  { 4 k }
      \Delta   m^2  \right )
       } {
       \sinh \left (  \dfrac {\pi L}{4k} ( k_\perp^2 + m _ {\rm h}^2 ) \right )
       \sinh \left (  \dfrac {\pi L}{4k} ( k_\perp^2 + m _ {\rm s}^2 ) \right )
       }
    \label{Msqu}
    .
\end{align}
Here one can explicitly see how $ | { \MM } | ^ 2 $ cuts off the
integrals over $ x $ and $ k _\perp $, both at small and at large
values. The result of these integrations will be a complicated
function of the thermal masses, and the result for the friction force will not be as
simple as without radiation \cite{Bodeker:2009qy}.
We find
\begin{align} \int d P _ {\rm split } \Delta  p _ { 1 \to 2 }
  =
  \frac { | C | ^ 2 g ^ 2 } { ( 2 \pi  ) ^ 3 }
  L ^{ -1 }
  \frac { \Delta  m ^ 2 } { m _ {\rm s } ^ 2 }
  f \left (
      \frac { \Delta  m  ^ 2 } { m _ {\rm s } ^ 2 }
      \right )
      \label{integral}
\end{align}
where the function $ f $ is given by the integral
\begin{align}
   f ( \xi  ) \equiv
   \int _ 0 ^ {\infty  } dz \dfrac { z  } { ( 1+z ) ^ 3 }
   \int _ 0 ^ {\infty  } du u
   \frac {
   \sinh \left ( \dfrac { u \xi  } { 1 + z } \right )
   } {
     \sinh ( u )
    \sinh \left ( u + \dfrac { u \xi  } { 1 + z } \right )
   }
   \label{f}
   .
\end{align}
For small and large $ \xi  $
\begin{align}
  f ( \xi  ) \simeq \frac { \pi  ^ 2 } { 36 } \xi  \qquad ( \xi  \ll 1 ),
  \qquad
  f(\xi) \to \frac{\pi^2}{24} \simeq 0.41 \qquad ( \xi  \to {\infty} )\,.
\end{align}
Thanks to this and to the
factor $ ( 2 \pi ) ^{ -3 } $ in (\ref{integral}) the numerical
prefactor in the splitting contribution to the friction is small.  The
splitting therefore only starts to dominate at quite large $\gamma$.

In the opposite limit of $mL \ll 1$, it is the WKB approximation which
breaks down.  In this case it is necessary to compute the full
$\chi(k)$ mode function in the presence of the wall, including the
finite reflection amplitude.  These mode functions were found
explicitly by Farrar and McIntosh \cite{Farrar:1994vp}, again assuming
a tanh wall profile.  We will not pursue this approach further here.

Finally we should point out another effect which can limit the
emission of the gauge bosons responsible for the friction.  Let us
estimate the phase-space density of the most important emitted gauge
bosons.  We found around \Eq{P_propto_gamma} that the pressure is
$\pres_{1\to 2} \sim \gamma g^2 m T^3$, arising from particles with
$k_\perp\sim k_z \sim m$.  Looking at \Eq{E_approx} we find the
force per particle is $p_z - k_z - q_z \sim m$.  So the density of
emitted particles is $n \sim \gamma g^2 T^3$.  This is to be compared
to the phase space $d^3 k \sim m^3$ they fill.  The mean occupancy is
\be
f(k) \sim \frac{n}{\Delta k^3} \sim \frac{g^2 \gamma T^3}{m^3} \,.
\ee
For large $\gamma > m^3 / g^4 T^3$, this can exceed the ``saturation''
occupancy $1/g^2$.  In this case, we expect that it is no longer safe
to consider different radiations as independent processes; there are
nonperturbatively large interactions between emitted quanta which
should suppress the emission process such that no phase space region
has occupancy higher than $\sim 1/g^2$.

With this in mind, let us rewrite \Eq{Msq_simp}, \Eq{phasespace} as
follows.  For the emission of transverse $W$ bosons, dropping order-1
constants but keeping the parametric dependence on $T,m,\gamma,g$,
the pressure is
\begin{align}
\nonumber
\pres_{1\to 2} & \sim \int d^2 k_\perp dk_z \Big( k^0 - k_z \Big)
\times
  \left[ \frac{g^2 k_\perp^2 m^4 \int f_p d^3 p}{k_z (k_\perp^2+m^2)^4}
    \right]
  \\ & \sim \int d^2 k_\perp dk_z \Big( k^0 - k_z \Big)
  \times
  \left[ \frac{g^2 m^4 \gamma T^3}{k_z k_\perp^6} \right] \,,
\label{saturate}
\end{align}
where $\int d^2 k_\perp dk_z$ is the phase space of emitted particles,
$(k^0-k_z)$ is the momentum transfer per particle,
and the quantity in square brackets is the occupancy
(phase space density) of emitted particles.  Roughly speaking,
saturation tells us to cut off the quantity in square brackets when it
exceeds $1/g^2$.  Doing so, the integral is dominated when
$k_z \sim k_\perp$.  The occupancy reaches $g^{-2}$ when
\begin{align}
k_z & \sim k_\perp \sim \left( \gamma g^4 T^3 m^4 \right)^{1/7} \,,
\\
   \pres_{1\to 2} & \sim \frac{k_\perp^4}{g^2} \sim \gamma^{\frac 47}
g^{\frac 27} T^{\frac{12}{7}} m^{\frac{16}{7}} \,.
\end{align}
This quantity rises with increasing $\gamma$ as $\gamma^{4/7}$, rather
than $\gamma^1$ as we found before.  Nevertheless, this is sufficient
to ensure that the pressure is enough to prevent a ``runaway'' wall,
$\gamma \ll 10^{10}$ so that the walls make up a negligible fraction
of the total energy in the Universe.

\section{Conclusions}
\label{sec:conclude}

In conclusion, if the plasma induces insufficient friction on the
bubble wall to prevent runaway, the wall accelerates to large
$\gamma$.  At leading order, the backwards friction on the bubble wall
from the medium scales as $\pres_{1\to 1} \sim m^2 T^2$, reaching a finite limit at
large $\gamma$.  The pressure driving the wall forward is generically
of the same scale, but may be numerically larger, in which case this
analysis alone would suggest a ``runaway'' situation with
$\gamma$ growing without bound.

But at the next order, transition radiation of
wall-frame soft, massive vector bosons introduces an additional
friction term with $\pres_{1\to 2} \sim \gamma g^2  m T^3$.  
Since this term rises
linearly with $\gamma$, it will limit the $\gamma$-factor of the wall
and prevent a true ``runaway.''  However, under criteria where the
leading-order calculation predicts runaway, we still expect a
parametrically large $\gamma$-factor, $\gamma \sim m/g^2 T$.  This
velocity is fast enough that, from the point of view of electroweak
baryogenesis, the wall can be considered to move with $v=1$.  For the
production of gravitational waves, one can treat $v \simeq 1$ but one
can neglect the energy accumulated by the bubble walls.

In some cases, the transition radiation may lead to a very high
occupancy of wall-frame soft $W,Z$ bosons, interacting
nonperturbatively with each other.  It is not clear to us what physics
might arise after the wall's passage, as these particles thermalize
with the broken-phase medium.  The occupancies are sufficient that
they could in principle generate sphaleron transitions immediately
after the passage of the wall.  We leave further considerations of
this problem to future investigations.

\section*{Acknowledgments}

The work of DB was supported in part by the Deutsche
Forschungsgemeinschaft under grant GRK881.  GM would like to thank the
TU Darmstadt and the Natural Science and Engineering Research Council
of Canada (NSERC) for partial funding of this work.  We are grateful
for conversations with Frank G\"ohmann, Thomas Konstandin, Miguel No,
Geraldine Servant, and Neil Turok (the latter conversation in 1994).

\bibliographystyle{unsrt}
\bibliography{split}

\end{document}